# Bibliometrics and Information Retrieval: Creating Knowledge through Research Synergies

**Judit Bar-Ilan**
Bar-Ilan University
Ramat Gan, Israel
Judit.Bar-Ilan@biu.ac.il

**Rob Koopman &
Shenghui Wang**
OCLC Research Europe
Leiden, Netherlands
Rob.Koopman@oclc.org,
shenghui.wang@gmail.com

**Andrea Scharnhorst**
Royal Netherlands Academy of
Arts and Sciences
Amsterdam, Netherlands
andrea.scharnhorst@dans.knaw.nl

**Marcus John**
Fraunhofer Institute
For Technological
Trend Analysis
Euskirchen, Germany
marcus.john@int.fraunhofer.de

**Philipp Mayr**
GESIS
Cologne, Germany
Philipp.Mayr-Schlegel@gesis.org

**Dietmar Wolfram**
University of Wisconsin-Milwaukee
Milwaukee, WI USA
dwolfram@uwm.edu

## ABSTRACT

This panel brings together experts in bibliometrics and information retrieval to discuss how each of these two important areas of information science can help to inform the research of the other. There is a growing body of literature that capitalizes on the synergies created by combining methodological approaches of each to solve research problems and practical issues related to how information is created, stored, organized, retrieved and used. The session will begin with an overview of the common threads that exist between IR and metrics, followed by a summary of findings from the BIR workshops and examples of research projects that combine aspects of each area to benefit IR or metrics research areas, including search results ranking, semantic indexing and visualization. The panel will conclude with an engaging discussion with the audience to identify future areas of research and collaboration.

## INTRODUCTION

Information Retrieval (IR) and Bibliometrics/Informetrics/Scientometrics (referred to hereafter as "metrics") represent two core areas of study in Information Science. Each has a long history with noted contributions to our understanding of how information is created, stored, organized, retrieved and used. Until recently, researchers have treated each of these areas as separate areas of investigation, with little overlap between the research topics undertaken in each area and little collaboration among researchers in both areas. This is surprising given that there are many common elements of interest to researchers in IR and metrics. Recognition of the mutually beneficial relationship that exists between IR and metrics has been growing over the past 15 years, with literature that specifically addresses this topic (e.g., Wolfram, 2003; Mayr & Scharnhorst, 2015) and the recent Bibliometric-enhanced Information Retrieval workshops (Mayr, Frommholz & Cabanac, 2016) held at metrics and IR meetings. The mutually beneficial relationship is evident in the application of metric and citation analysis methods in the design of IR systems and in the use of techniques developed in IR that lend themselves to the study of metric phenomena. A prime example is the development and use of the PageRank algorithm by Page, Brin, Motwani and Winograd (1999), which was inspired by ideas from citation analysis and then adapted to the Web to inform relevance ranking decisions of documents. It has since been re-purposed by metrics researchers for the ranking of authors and papers.

**PANEL ORGANIZATION**

This panel brings together researchers in IR and metrics to present an overview of how IR and metrics research may be combined, to provide examples of research that intersect both areas, and to engage in a discussion with the audience about future potential topics. The session will begin with an overview of the synergies that exist between IR and metrics, followed by a summary of findings from the BIR workshops and examples of research projects that combine aspects of each area to benefit IR and/or metrics research. The panel will conclude with an engaging discussion with the audience to identify future areas of research and collaboration. Initial questions to stimulate the discussion will include: 1) why don't more IR researchers look to metrics research to help solve their research problems and vice versa, and; 2) as an IR or metrics researcher, what do you see as a research problem of interest that could benefit from the approaches used by the other area?

**OVERVIEW (Dietmar Wolfram)**

Metrics researchers have long recognized that empirical regularities or patterns exist in the way information is produced and used, such as author and journal productivity, the way language is used, and how literatures grow over time. These regularities extend to the content of IR systems and how the systems are used. Knowledge of these regularities, such as patterns in how users interact with IR systems, can help to inform the design and evaluation of IR systems. Similarly, measures developed for metrics research also have applications in IR.

Conversely, techniques developed that support more efficient IR are now being applied in metrics studies. This is exemplified in the use of language and topic modeling, which were developed to overcome limitations of more simplistic "bag of words" approaches in IR. Topic modeling has become a useful tool for better understanding relationships between papers, authors and journals by relying on the language used within the documents of interest. These tools complement existing methods based on citations and collaborations in helping researchers to reveal the underlying structure of disciplines. The relationships between metrics research and IR--and in particular academic search--are much closer than many researchers realize.

**RECENT ADVANCES IN BIBLIOMETRIC-ENHANCED INFORMATION RETRIEVAL (Philipp Mayr)**

The presentation will report about recent advances of the Bibliometric-enhanced Information Retrieval (BIR) workshop initiative.

Our motivation as organizers of the BIR workshops (2014, 2015 and 2016) started from the observation that the main discourses in both fields are different and the communities only partly overlap, as well as from the belief that a knowledge transfer is profitable for both sides.

The first BIR workshop in 2014 set the research agenda by introducing each group to the other, illustrating state-of-the-art methods, reporting on current research problems, and brainstorming about common interests. The second workshop in 2015 further elaborated these themes. The third full-day BIR workshop at ECIR 2016 aimed to establish a common ground for the incorporation of bibliometric-enhanced services into scholarly search engine interfaces. In particular, we addressed specific communities, as well as studies on large, cross-domain collections like Mendeley and ResearchGate. The third BIR workshop addressed explicitly both scholarly and industrial researchers. In June 2016, we will organize the 4th BIR workshop at the JCDL conference in collaboration with the NLP and computational linguistics research group from Min-Yen Kan (see BIRNDL workshop http://wing.comp.nus.edu.sg/birndl-jcdl2016/).

The past workshop topics included (but were not limited to) the following:

- IR for digital libraries and scientific information portals
- IR for scientific domains, e.g. social sciences, life sciences, etc.
- Information seeking behavior
- Bibliometrics, citation analysis, and network analysis for IR
- Query expansion and relevance feedback approaches
- Science Modeling (both formal and empirical)
- Task-based user modeling, interaction, and personalization
- (Long-term) Evaluation methods and test collection design
- Collaborative information handling and information sharing
- Classification, categorization, and clustering approaches
- Information extraction (including topic detection, entity and relation extraction)
- Recommendations based on explicit and implicit user feedback

Previous BIR workshops have generated a wide range of papers. Proceedings are available at http://ceur-ws.org/Vol-1143/, http://ceur-ws.org/Vol-1344/ and http://ceur-ws.org/Vol-1567/. The main directions of these workshop papers have been:

- IR and recommendation tool development and evaluation
- Bibliometric IR experiments and data sets
- Document Clustering for IR
- Citation Contexts and Analysis

The presentation will report about highlights of the past workshop papers and outline future directions of this initiative.

**APPLICATION OF THE H-INDEX FOR RANKING SEARCH RESULTS (Judit Bar-Ilan)**

In traditional IR, search results are usually ranked using tf*idf (term frequency/inverse document frequency). On the web, hypertext links can be utilized as well. The web-graph (node=web pages, links=hypertext links) is similar to citation networks (nodes=publications, links=citations). Citations are usually counted without assigning weights to citation. Similarly, the number of links to a web page can be counted, but this turns out to be insufficient because of the lack of quality control on the web, and links have to be weighted by their "importance". This is the idea behind the PageRank (Page et al., 1999). This idea stems from bibliometrics (Pinski & Narin, 1976). It should be noted that the PageRank calculation is quite costly.

We suggest using a variant of the h-index for ranking. The h-index was introduced by Jorge Hirsch (2005). Hirsch is a physicist, but the idea of combining publication and citation counts captured the imagination of bibliometrics researchers, and a huge number of variants were suggested. One of them, the h-index of a single journal paper suggested by Schubert (2009), can be applied to the web graph as well, by assessing the importance of a web page by the number of inlinks webpages linking to this page received (Bar-Ilan & Levene, 2015). The advantage of this method is that it is based on local computation unlike PageRank. This idea shows how bibliometrics and information retrieval can inform each other.

**SEMANTIC INDEXING FOR INFORMATION RETRIEVAL AND BIBLIOMETRIC ANALYSIS (Rob Koopman & Shenghui Wang)**

Large scale digital libraries offer users the opportunities to explore a vast amount of information using relatively uniform mechanisms, such as keyword-based or faceted searches. In the meantime, users are challenged to make sense of the overloaded result sets that are too big and complex to comprehend or to understand and counteract the biases derived from different ranking mechanisms that render the results. We believe that semantic indexing based on statistical analysis together with intuitive interfaces can help users to find relevant information and discover patterns fast and reliably.

In this talk, we will present our Ariadne context explorer, which allows users to visually explore the context of bibliographic entities, such as authors, subjects, journals, citations, publishers, etc. The visualization is built on semantic indexing of these entities based on the terms that share the same contexts in a large scale bibliographic dataset. The statistical analysis based on Random Projection results in an underlying semantic space within which each entity is represented vectorially. Each bibliographic record or any piece of text could also be represented as a vector in this semantic space. The information retrieval task then becomes a task of finding the nearest neighbors in this space, no matter the search starts with an author, a citation, an article or a free text.

We will demonstrate the Ariadne context explorer and report the results of applying such semantic indexing and visualization in a topic-delineation exercise.

**SEEKING FOR THE NEEDLE IN THE HAYSTACK: BIBLIOMETRICS, INFORMATION RETRIEVAL AND VISUALIZATION IN THE CONTEXT OF TECHNOLOGY FORESIGHT (Marcus John)**

Technology foresight is an important element of any strategic planning process, since it assists decision makers in identifying and assessing future technologies. One important assumption made in this context is that tomorrow's technologies are based on today's daily work in scientific laboratories. Consequently, any technology foresight process must rely on a continuous scanning of the scientific and technological landscape in order to detect scientific advances, breakthroughs and emerging topics. In other words, a kind of science observatory has to be established. Due to the rising number of scientific papers published each year, it becomes more and more difficult to restrict this scanning and monitoring process solely to classical desktop research and information retrieval techniques. Additionally the classical task of IR, namely the identification of relevant information is exacerbated by the need to identify relevant and new information. Consequently, the information overload makes it necessary to complement classical approaches by quantitative data-driven approaches stemming from informetrics, bibliometrics, data mining and related fields.

This work in progress report presents an overview of the ongoing research at the Fraunhofer INT and addresses the question if and how these quantitative data-driven approaches might enhance the classic portfolio of technology foresight. This will be exemplified along a prototypical technology foresight process along which different IR-related challenges will be identified. It will be demonstrated how eavesdropping into today's scientific communication by bibliometric means might support this process. Exemplarily, a procedure coined "trend archaeology" will be presented. This approach examines historic scientific trends and seeks for specific patterns within their temporal evolution. The proposed method is a multidimensional approach, since it takes into account multiple aspects of a scientific theme using bibliometric means. Additionally, "trend archaeology" is based on the synoptic inspection of different scientific themes, which emanate from different fields like nanotechnology or materials science. It will be demonstrated that for technology foresight it is mandatory to take into account the multidimensional-multiscalar, dynamic and highly interconnected nature of science.

**THE PANEL MEMBERS**

**Andrea Scharnhorst** (moderator) is Head of e-Research at the Data Archiving and Networked Services (DANS) institution in the Netherlands - a large digital archive for research data primarily from the social sciences and humanities. She is also member of the e-humanities group at the Royal Netherlands Academy of Arts and Sciences (KNAW) in Amsterdam, where she coordinates the computational humanities programme. Her work focuses on understanding, modeling and simulating the emergence of innovations.

**Judit Bar-Ilan** (panelist) is professor at the Department of Information Science of Bar-Ilan University in Israel. She received her PhD in computer science from the Hebrew University of Jerusalem and started her research in information science in the mid-1990s at the School of Library, Archive and Information Studies of the Hebrew University of Jerusalem. She moved to the Department of Information Science at Bar-Ilan University in 2002. Her areas of interest include informetrics, information retrieval, Internet research, information behavior, the semantic Web and usability. Additional details are available at: http://is.biu.ac.il/en/judit/.

**Marcus John** (panelist) received his PhD in the field of theoretical astrophysics. Since 2007, he has been a senior scientist at the Fraunhofer Institute for Technological Trend Analysis where he is mainly concerned with technology foresight and future-oriented technology analysis. His main fields of interest are complex systems science, physics of socio-economic systems, simulation methods and human enhancement. Additionally his work focuses on the application of bibliometric and other quantitative methods for technology foresight.

**Rob Koopman** (panelist) is an architect in OCLC European, Middle East and Africa (EMEA) office in Leiden, Netherlands. His main research area is applied data science. He has a physics background and has worked at OCLC EMEA since 1981.

**Philipp Mayr** (panelist) is team leader at the GESIS department Knowledge Technologies for the Social Sciences (WTS). He is the main organizer of the past BIR workshops. His team at GESIS runs two retrieval platforms which cover bibliographic information and full texts in the social sciences. His research interests include informetrics, information retrieval and digital libraries. Additional details are available at: http://www.gesis.org/de/das-institut/mitarbeiterverzeichnis/?alpha=M&name=philipp%2Cmayr.

**Shenghui Wang** (panelist) is a research scientist in OCLC Research since 2012, based in OCLC EMEA office in Leiden, Netherlands. Her current research activities include text mining, visualization as well as linked data investigations.

**Dietmar Wolfram** (panelist) is professor at the School of Information Studies at the University of Wisconsin-Milwaukee. He received his PhD in Library and Information Science from the University of Western Ontario. His research interests include informetrics, information retrieval, the intersection between these two areas, scholarly communication and user studies.

**ACKNOWLEDGMENTS**

**REFERENCES**

Bar-Ilan, J., & Levene, M. (2015). The hw-rank: An h-index variant for ranking web pages. *Scientometrics,* 102, 2247-2253.

Hirsch, J. E. (2005). An index to quantify an individual's scientific research output. *Proceedings of the National academy of Sciences of the United States of America*, *102*(46), 16569-16572.

Mayr, P., Frommholz, I., & Cabanac, G. (2016). Bibliometric-Enhanced Information Retrieval: 3rd International BIR Workshop. In N. Ferro et al. (Eds.), *Advances in Information Retrieval: 38th European Conference on IR Research, ECIR 2016* (pp. 865-868). Springer.

Mayr, P., & Scharnhorst, A. (2015). Scientometrics and information retrieval - weak-links revitalized. Scientometrics, 102(3), 2193–2199. doi:10.1007/s11192-014-1484-3

Page, L., Brin, S., Motwani, R., & Winograd, T. (1999). The PageRank citation ranking: bringing order to the web. URL: http://ilpubs.stanford.edu:8090/422/1/1999-66.pdf.

Pinski, G., & Narin, F. (1976). Citation influence for journal aggregates of scientific publications: Theory, with application to the literature of physics. *Information Processing and Management*, 12(5), 297–312.

Schubert, A. (2009). Using the h-index for assessing single publications. *Scientometrics*, 78(3), 559–565.

Wolfram, D. (2003). *Applied informetrics for information retrieval research*. Westport, CT: Libraries Unlimited.